\pdfoutput=1 

\documentclass[sigplan,screen,10pt,authorversion]{acmart}
\usepackage{latexsym}
\usepackage{multirow}
\usepackage{amsmath}

\usepackage{color}
\definecolor{keywordcolor}{rgb}{0.7, 0.1, 0.1}   
\definecolor{tacticcolor}{rgb}{0.0, 0.1, 0.6}    
\definecolor{commentcolor}{rgb}{0.4, 0.4, 0.4}   
\definecolor{symbolcolor}{rgb}{0.0, 0.1, 0.6}    
\definecolor{sortcolor}{rgb}{0.1, 0.5, 0.1}      
\definecolor{attributecolor}{rgb}{0.7, 0.1, 0.1} 

\usepackage[normalem]{ulem}

\title{Trustworthy Formal Natural Language Specifications} 
\author{Colin S. Gordon}
\orcid{0000-0002-9012-4490}
\email{csgordon@drexel.edu}
\affiliation{%
  \institution{Drexel University}
  \city{}
  \country{USA}
}

\author{Sergey Matskevich}
\orcid{0000-0002-8080-7298}
\email{sm3372@drexel.edu}
\affiliation{%
  \institution{Drexel University}
  \city{}
  \country{USA}
}

\usepackage{mathpartir}
\usepackage{graphicx}
\usepackage{xcolor}
\usepackage{listings}
\lstdefinelanguage{scala}{
  morekeywords={abstract,case,catch,class,def,%
    do,else,extends,false,final,finally,%
    for,if,implicit,import,match,mixin,%
    new,null,object,override,package,%
    private,protected,requires,return,sealed,%
    super,this,throw,trait,true,try,%
    type,val,var,while,with,yield},
  otherkeywords={=>,<-,<\%,<:,>:,\#,@},
  sensitive=true,
  morecomment=[l]{//},
  morecomment=[n]{/*}{*/},
  morestring=[b]",
  morestring=[b]',
  morestring=[b]"""
}
\usepackage[T1]{fontenc}
\usepackage[utf8]{inputenc}

\lstdefinelanguage{Coq}{
    morekeywords={Proof,Qed,Polymorphic,Lemma,Hint,Unfold,Local,Obligation,Tactic,Ltac,Instance,Class,Defined,Require,Import,Section,End,Parameter,Record,Program,Example,Inductive,Infix,Notation,Definition,Fixpoint,Postulate,Goal},
    morekeywords=[2]{eauto},
    keywordstyle=[2]\color{teal},
    morekeywords=[3]{Set,Prop,Type,with,forall,using,fun},
    keywordstyle=[3]\color{olive},
    otherkeywords={->,|,:,=},
    sensitive=true,
    morecomment=[n]{(*}{*)},
    basicstyle=\small\ttfamily,
    keywordstyle=\color{blue},
    commentstyle=\color{violet}\small\ttfamily}
\usepackage{xspace}
\usepackage{stmaryrd}
\usepackage{amsmath}

\newcommand{\coq}{\textsc{Coq}\xspace}
\newcommand{\lean}{\textsc{Lean}\xspace}
\newcommand{\cic}{\textsc{CIC}\xspace}

\setcopyright{acmlicensed}
\acmPrice{15.00}
\acmDOI{10.1145/3622758.3622890}
\acmYear{2023}
\copyrightyear{2023}
\acmSubmissionID{onward23papers-p45-p}
\acmISBN{979-8-4007-0388-1/23/10}
\acmConference[Onward! '23]{Proceedings of the 2023 ACM SIGPLAN International Symposium on New Ideas, New Paradigms, and Reflections on Programming and Software}{October 25--27, 2023}{Cascais, Portugal}
\acmBooktitle{Proceedings of the 2023 ACM SIGPLAN International Symposium on New Ideas, New Paradigms, and Reflections on Programming and Software (Onward! '23), October 25--27, 2023, Cascais, Portugal}
\received{2023-04-28}
\received[accepted]{2023-08-11}

\begin{document}
%

\begin{abstract}
  Interactive proof assistants are computer programs carefully constructed to check a human-designed proof of a mathematical claim with high confidence
  in the implementation.
  However, this only validates truth of a formal claim, which may have been mistranslated from a claim made in natural language.  
  This is especially problematic when using proof assistants to formally verify the correctness of software with respect to a natural language 
  specification.
  The translation from informal to formal remains a challenging, time-consuming process that is difficult to audit for correctness.
  \looseness=-1

  This paper shows that it is possible to build support for specifications
  written in expressive subsets of natural language,
   within existing proof assistants, consistent with 
  the principles used to establish trust and auditability in proof assistants themselves.
  We implement a means to provide specifications in a modularly extensible formal subset of English, 
  and have them automatically translated into formal claims, entirely within the 
  Lean proof assistant. Our approach is extensible (placing no permanent restrictions on grammatical structure),
  modular (allowing information about new words to be distributed alongside libraries), 
  and produces proof certificates explaining how each word was interpreted and how the sentence's structure was used to compute the meaning. 

  We apply our prototype to the translation of various English descriptions of formal specifications from a popular textbook into Lean formalizations;
  all can be translated correctly with a modest lexicon with only minor modifications related to lexicon size.
\end{abstract}
\maketitle

\section{Introduction}
Proof assistants can establish very high confidence in the correctness of formal proofs, due to both their rigorous checking and attention to producing independently auditable evidence that the arument is correct~\cite{paulson1990logic,pollack1998believe}.
But one of the unavoidable points of trust for even a carefully-implemented proof assistant is the specifications themselves: proving the wrong theorem is of limited use. And only those who can read both formal and informal specifications can even consider whether this has occurred. This is particularly crucial for software verification: 
software specifications typically originate in natural language, and any accompanying formal specification comes afterwards --- which increasingly occurs for compilers~\cite{leroy2009formally}, operating systems~\cite{seL4TOCS}, and other high-value software.
Currently, the \emph{only} bridge between the formal and informal specifications is the humans who perform the translation. 
There is no independently checkable record of this translation aside from the possibility of comments or notes by the translators --- themselves largely in informal (though likely meticulous) natural language.
Simply being familiar with both the specification language and the intended specification is insufficient by itself to bridge this gap~\cite{finney1996empirical,finney1996mathematical}: relating the two is a separate skill that is independently challenging to develop.

Ideally, it would be possible to give natural language specifications directly to the proof assistant, for example:
\begin{lstlisting}[language=lean]
theorem thm : "addone is monotone" := ...
\end{lstlisting}
Robust support for such specifications could enable significant improvements in requirements tracing (machine \emph{checked} mappings from natural language to formal results), including for artifact evaluation; education, where it could help students check their understanding of how either mathematics or program specifications are formalized logically; and even communication with non-technical clients who might wish to have some confidence that a formalization they do not themselves fully understand is correct. On the last point, Wing's classic paper introducing formal methods~\cite{wing1990specifier} posits that customers may read the formal specifications produced from informal requirements, but this is only possible if the client can make sense of the formal specification itself. Machine-checked relationships between natural-language and formalized expressions of software properties can help bridge this knowledge gap by connecting formal properties to natural language a reader with less background in a specific formal logic could understand.

This paper develops the core of an \emph{extensible} approach to allow specifications like:
\begin{lstlisting}
theorem thm : pspec [| addone is monotone |] := ...
\end{lstlisting}
(where \lstinline|pspec| returns the logical form of the natural language utterance in quotes).
We envision such a system can eventually be used for the purposes above, to generate formal claims about mathematics or programs verified in a proof assistant, whether specified in a proof assistant's own logic, or indirectly through a foundational program logic~\cite{appel2001foundational} built inside a proof assistant; we primarily describe an implemention in an early version of \lean 4, with some discussion of a parallel prototype in \coq.
Our goal is to enable 
specification of formal claims using modularly-extensible fragments of natural language,
within the proof assistant, which are automatically translated to forms logically equivalent (and in practice often near-identical) 
to those a verification expert may give while accounting for what is easier or more difficult to prove; 
proofs (often trivial) relating the expert claim to the formalization extracted from natural language then 
extend the formal proof with a formalized connection to a description of functionality that may be more easily understood by non-experts in verification, 
additionally ensuring that the mathematical formalization is consistent with a less formal statement.


This is not a job for machine-learning-centric natural language processing, which despite the recent trend in work on ``autoformalization''~\cite{llm-autoformalization} is incompatible with the goals of using a proof assistant for formal verification. 
There is no guarantee a learned translation is sensible, and if a translation from natural to formal language ends up being surprising, few machine learning approaches produce an auditable trail of evidence for \emph{why} that translation was produced by the trained model, and no way to precisely fix misunderstandings of specific words. The current state-of-the-art in explaining such translations are techniques such as \emph{salience maps}, which essentially highlight which input words were most influential to the system output without providing a linguistically-grounded reason for the behavior~\cite{sogaard2021explainable}.
Meanwhile, \emph{proof certificates} play a central role in the design of trustworthy proof assistants~\cite{pollack1998believe,gordon2000lcf} and foundational program verification~\cite{appel2001foundational}.
Moreover, as proof assistants are often used to formalize properties of new mathematics or new programs, often using new terminology, there will often be a lack of training data for mapping natural language to a formal property.
Later, we point out additional ways that the needs of trusted formalization of natural language specifications run afoul of many of machine learning's known limitations, while requiring few of its advantages. (We also point out limited ways machine learning can play a role in optimizing the techniques we employ.)

Fortunately the field of linguistics predates machine learning. Formalizing \emph{categorial grammar}~\cite{ajdukiewicz1935syntaktische,bar1953quasi,lambek1958mathematics,Steedman:2001} carefully in a proof assistant offers a path to a natural, auditable way to bridge the gap between formal and natural-language specification.
In this paper we show a prototype demonstrating that it is possible to parse a string containing a natural language specification into a semantic representation that can be used directly in proofs within a proof assistant (i.e., a proposition in the Calculus of Inductive Constructions), in a linguistically principled way, using typeclasses~\cite{sozeau2008first,selsam2020tabled}.
We argue that this approach is modular and can extend to sophisticated and substantial subsets of natural language.
We also analyze how the trusted computing base is affected when considering trust of a formal verification up to the natural language specification.
Unlike many previous attempts at natural language interfaces (Section \ref{sec:relwork_natspec}), our goal is to supplement, not supplant, traditional
proof assistant interfaces.

This paper establishes that the theoretical core is within reach, 
that these techniques can capture the grammar and semantics
of a small sample of natural language specifications  from a textbook on verified functional programs,
and that it is feasible to implement translation algorithms inside
current proof assistants, as a library (though leading to suggestions for improvements to those
features).

  \section{Background \& Motivation}\label{background}
  \begin{figure*}
  \begin{mathpar}
    \inferrule[$\backslash$-Elim]{\Gamma\vdash A \Rightarrow a\quad \Delta\vdash A\backslash B \Rightarrow f}{\Gamma,\Delta\vdash B \Rightarrow (f~a)}
    \quad
    \inferrule[/-Elim]{\Gamma\vdash A/B \Rightarrow f\quad \Delta\vdash B \Rightarrow a }{\Gamma,\Delta\vdash A \Rightarrow (f~a)}
    \quad
  \inferrule[/-Comp]{\Gamma\vdash A/B \Rightarrow e \quad \Delta \vdash B/C \Rightarrow f }{\Gamma,\Delta\vdash A/C \Rightarrow (e\circ f)}
  \quad
  \inferrule[$\backslash$-Comp]{\Gamma\vdash A\backslash B \Rightarrow e \quad \Delta \vdash B\backslash C \Rightarrow f }{\Gamma,\Delta\vdash A\backslash C \Rightarrow (f\circ e)}
  \end{mathpar}
  \vspace{-1.5em}
  \caption{A selection of rules used in this paper, all derivable in \textsc{CTL}s, \textsc{CCG}, and other categorial grammars.
  }
  \label{fig:rules}
  \end{figure*}

  This section provides a condensed (and therefore somewhat biased)
  background in natural language processing and categorial grammar, from a programming languages 
  perspective.
  
  \subsection{Categorial Grammars}
  \emph{Categorial grammars} are a family of related techniques~\cite{oehrle1988categorial} applying ideas from logic
   to relate the syntax of natural language 
  with formal representations of what that syntax denotes --- often called \emph{logical form} --- growing out of Lambek's work~\cite{lambek1958mathematics}
  in the 1950s (which coined the term).
  The core idea is to build a sort of substructural type theory where base types correspond to 
  grammatical \emph{categories} (hence \emph{categorial}), from which more complex grammatical 
  categories can be defined.  A set of inference rules is then used to define, simultaneously, 
  how grammatical categories combine into larger sentence fragments and how those smaller fragments' 
  meanings (logical forms) are combined into larger meanings. This process bottoms-out at a \emph{lexicon}, 
  giving for each word its grammatical roles (types) and associated denotations. Thus categorial grammar is 
  a system of simulatenously parsing natural language from strings and assigning denotational semantics 
  --- a process traditionally referred to as \emph{semantic parsing} in the computational linguistics 
  literature.

  Most prominent in the linguistics community are \emph{combinatory categorial grammars} 
  (\textsc{CCG}s)~\cite{Steedman:2001}, though also relevant are the 
  \emph{categorial type logics} (\textsc{CTL}s~\cite{carpenter1997type}\footnote{Occasionally also called \emph{type-logical grammars} (\textsc{TLG}s).}).  Work on \textsc{CCG}s epmhasizes appropriate constructs for linguistic ends, while \textsc{CTL}s hew close to Lambek's view~\cite{lambek1958mathematics} of categorial grammars as substructural logics for linguistics. 
  While these reflect very different philosophical and practical aims, for our present purposes the distinction is immaterial: it is widely held, and in some cases formalized~\cite{kruijff2000relating,gerhard2005anaphora}, that rules used in \textsc{CCG}s (including the variant with the most sophisticated linguistic treatments~\cite{Baldridge:2003:MCC:1067807.1067836}) correspond to theorems in particular \textsc{CTL}s~\cite{Moortgat1996}.
  In this work we use only principles common across all categorial grammars, including
   \textsc{CCG}s and \textsc{CTL}s.

  
  All categorial grammars parse by combining sentence fragments based on their grammatical types.  
  These types include both atomic primitives (such as noun phrases) as well as more complex types, 
  called \emph{slash-types} that indicate a predicate argument structure (which are used to model, 
  for example, most classes of verbs).
  Oversimplifying slightly, categorial grammars treat parsing as logical deduction
   in a residuated non-commutative linear logic.\footnote{Technically only \textsc{CTL}s~\cite{morrill2012type} take this as an epistemilogical commitment, while \textsc{CCG}s~\cite{Steedman:2001}) are agnostic, inheriting such a relation via Baldridge and Kruijff's work~\cite{Baldridge:2003:MCC:1067807.1067836}.} This is essentially a family of linear logics without the structural rule for freely commuting the order of assumptions, thus modeling sensitivity to word order, and picking up as a consequence \emph{two} forms of implication corresponding to whether an implication expects its argument to the left or to the right.\footnote{It is the presence of the ability to commute assumptions arbitrarily that allows a single implication to suffice in standard logics.}
  The model for the logic is a sequence of words, and types correspond to
  the grammatical role of a sentence fragment. 

  $A/B$ is the grammatical type for a fragment that, when given a $B$ to its right, forms an $A$.  
  $A\backslash B$ is the grammatical type for a fragment that, when given an $A$ to its left, forms a $B$.  
  In both cases, the argument is ``under'' the slash, and the result is ``above'' 
  it.\footnote{We follow CTL notation rather than CCG notation (which always puts results to the left) as users of proof assistants tend to be familiar with a range of logics, so the CCG syntax would likely confuse users already familiar with the Lambek calculus and related systems. This notational choice is orthogonal to the choice of which rules to employ.})
  These are called \emph{slash types}.
  The grammars include rules to combine adjacent parts of a sentence.
  The elimination rules for slash types are the first two in Figure \ref{fig:rules}.
  The judgment $\Gamma\vdash C \Rightarrow e$ is read as claiming the sequence of words $\Gamma$ can be combined to form a 
  sentence fragment of grammatical type $C$, whose underlying semantic form --- logical form --- is given by $e$.  
  $e$ is a term drawn from the logical language being used to represent sentence meaning, typically a simply-typed lambda 
  calculus in keeping with Montague~\cite{montague1970universal,montagueEFL70,montaguePTQ73,partee1997montague}, though in 
  our work we follow the alternative~\cite{bekki14dts,sundholm1986proof,chatzikyriakidis2017modern,ranta1994ttg} 
  of targeting a dependently-typed calculus.
  Figure \ref{fig:rules} also includes forward and backward composition; rules tend to come in pairs for each direction.
  The rest of the paper can be followed with only the four rules of this figure, though our implementation includes additional
  rules from the formal linguistics literature.

  A \emph{lexicon} gives the grammatical role and semantics for individual words, providing the starting point for combining fragments.
  Categorial grammars push all knowledge specific to a particular human language into the lexicon, 
in categorizing how individual words are used. This allows the core principles to be reused across languages,
as evidenced by wide-coverage lexicons for a variety of natural 
languages including English~\cite{hockenmaier2007ccgbank}, German~\cite{hockenmaier2006creating},
  Hindi~\cite{ambati2018hindi}, Japanese~\cite{mineshima2016building}, Arabic~\cite{boxwell2010pilot}, French~\cite{moot2015type}, and Dutch and Italian~\cite{abzianidze2017parallel}.
  
\begin{figure*}
\begin{mathpar}
\inferrule*[right=$\backslash$-Elim]{
    \inferrule*{ }{\textrm{``four''}\vdash NP \Rightarrow 4}\;
    \inferrule*[right=/-Elim]{
      \inferrule{ }{\textrm{``is''} \vdash (NP\backslash S)/ADJ \Rightarrow \lambda p\ldotp\lambda x\ldotp p~x}\;
      \inferrule{ }{\textrm{``even''} \vdash ADJ \Rightarrow \mathsf{even}}
    }{\textrm{``is~even''}\vdash NP\backslash S\Rightarrow (\lambda x\ldotp \mathsf{even}~x)}
}{
    \textrm{``four~is~even''}\vdash S\Rightarrow \mathsf{even}~4
}
\end{mathpar}
\vspace{-2em}
\caption{A simple semantic parse using categorial grammar.}
\label{fig:example}
\end{figure*}
  Together, these allow filling in choices for the metavariables in the rules above, permitting derivations like that in Figure \ref{fig:example}.
  Each grammatical type $C$ corresponds to a particular type in the underlying lambda calculus and the underlying \emph{semantic} type is determined by a systematic translation from the syntactic grammatical type.
  Borrowing more notation from logic (where this idea is known as a Tarski-style universe~\cite{martin1984intuitionistic}), we write $\lfloor C\rfloor$ for $C$'s semantic type.  
  Both slash types correspond to function types in the lambda calculus: 
  $ \lfloor A/B\rfloor = \lfloor B\backslash A\rfloor = B\rightarrow A$.
  An invariant of the judgment $\Gamma\vdash C\Rightarrow e$ is that in the underlying logic, $e$ always has type $\lfloor C\rfloor$.
  This invariant explains why it is correct for ``four'' to have semantics $4$ while ``is even'' has a function as its semantics.
    In proof assistants based on type theory, like \coq and \lean, the set of grammatical types can be given as a datatype declaration, and the interpretation function as a function from grammatical types to proof assistant types.

  These few rules are enough to formalize a small fragment of English in ways consistent with
  linguistics research, and demonstrate the possibility of interpreting natural language specifications \emph{within} a proof assistant, in a well-founded and extensible way.  Our initial choice of rules is limited, but not fundamentally so.
   \textsc{CCG}s recognize \emph{mildly context-sensitive languages}~\cite{joshi:91,Kuhlmann2015}, which are believed to cover
   the full range of grammatical constructions in any natural language~\cite{steedman2012taking}. 
  All rules of \textsc{CCG}s are encodable in the way we describe, as are the rules of 
  Turing-complete \textsc{CTL}s~\cite{carpenter1999turing}. Ultimately the question of which rules are required is an 
  empirical one, considering both linguistic constructions and the complexity of recognizing these grammars
  For this paper we focus on a small set of rules including those above, which are common to most categorial grammars.
  The rules we use in this paper are intended as a foundation: as the next section points out, categorial grammars are highly modular,
  allowing not only simple additions of new lexical entries, but simple modular addition of support for new grammatical constructions,
  often by direct implementation of treatments that have been extensively validated in the linguistics literature. 

  \subsection{Formal Grammars for Natural Languages: Why Categorial Grammars?}
  \label{sec:formal_natural}
  Producing formal grammars (categorial or otherwise) for natural language is technically a slight misnomer.
  Natural language is inherently open-ended, allowing new grammatical constructions to become accepted over time,
  and new words to be proposed, among other changes; we are ignoring, for present purposes, matters of pronunciation 
  (phonology and phonetics) and changes in written form (orthography), not to mention signed languages.
  However, as soon as a set of acceptable words and their uses are fixed into a formal grammar, that grammar now describes
  merely a subset of the actual natural language it models, and by its nature excludes innovations in the lexicon and grammatical structure.
  This has far-reaching effects if our interest is in writing formal specifications in something
  approximating natural language: individual programming projects often invent their own terminology!
  Thus if we use categorial grammars (or any other formal theory of grammar) to parse specifications into a logical form,
  we are necessarily writing formal specifications in a \emph{formalized subset} of natural language,\footnote{Hence the double meaning of our title.}
  rather than open-ended natural language.

  This is an inherent limitation to any approach to mechanized manipulation of language meaning.\footnote{As opposed to mechanized manipulation of language
  \emph{form} (text) alone, as is done in current large language models, where open-endedness is well-supported (e.g., via the well-known \emph{attention}~\cite{vaswani2017attention}
  mechanism, which essentially heuristically identifies chunks of input text to copy into the output), but meaning is not (which is why large language models consistently
  struggle with basic boolean reasoning~\cite{ettinger2020bert,pandia-ettinger-2021-sorting,traylor2021and,truong-etal-2023-language}).}
  The result is that novel grammatical constructions (or simply some not accounted for in the design of the grammar),
  or previously-unknown words will not be handled \emph{at all} by the formal grammar.
  This is why the approach we are suggesting is sometimes called \emph{controlled natural language}~\cite{fuchs1999controlled,kuhn2014survey}.

  However, choice of grammar formalism can moderate these limitations. Rather than attempting to treat all possible current
  and future natural language sentences out of the box (which is impossible), we can choose foundations that are modularly extensible
  to new words \emph{and new grammatical constructions},
  and can grow incrementally over time as terminology or grammar evolves or changes.
  Categorial gramamrs are examples of \emph{lexicalized} grammar formalisms~\cite{schabes:90}. In contrast to the context-free grammars
  that are well-known in programming languages and software engineering, all knowledge of how an individual word is used is contained in the lexicon alone.
  Whereas a CFG treatment of English might have separate non-terminals for verb phrases, transitive verbs, and intransitive verbs, and individual words would simply be terminals,
  in categorial grammars all verbs, including ditransitive verbs (those with direct and indirect objects) are simply built from noun phrases and slash types.
  A good example of a non-trivial modular addition in our current prototype is in-situ quantification: when a use of the indefinite article ``a''
  in the \emph{middle} of a sentence introduces a top-level existential quantifier in the semantics, as in ``addone given 3 returns a natural''
  --- this translates to $\exists x:\mathbb{N}\ldotp \mathsf{addone}(3)=x$, but knowledge of the existential handling is isolated to the entry for ``a.''
  Adding this support to other systems (Section \ref{sec:relwork_natspec}) is non-trivial.

  Because categorial grammars focus on a set of base categories plus categories that correspond to functions (slash types), the modularity
  is improved even beyond other lexicalized formalisms, which require a finite set of grammatical categories.
  Additional slash-based categories can be introduced freely.
  And the menagerie of categorial grammar variants are largely completely compatible with each other: the \emph{rules} for categorial
  grammar are modular. 
  This is a suitable basis for tackling a problem whose true empirical needs may not be fully established for years to come, allowing the system
  to evolve in a flexible way.

  Categorial grammars have been extensively validated for modeling a wide array of complex grammatical
phenomena across a diverse array of language families, and
the linguistics literature contains decades' worth of work studying how to
capture subtle linguistic semantic phenomena in these languages via categorial 
grammars~\cite{carpenter1997type,morrill2012type,jacobson2014compositional,barker2014continuations,abzianidze2017parallel,moot2012logic}.
This means that building on (for now) basic categorial grammars is forward-compatible with decades of thoroughly-validated
work on detailed linguistic theories accounting for both syntax \emph{and} semantics of significant natural language fragments.
If it is discovered that an additional grammatical structure is necessary or useful, we can essentially
consult the linguistics literature and add just the required rules;
it is  long-standing practice for the different branches
to borrow each others' innovations; with rare exception, these combinations are straightforward and require almost no adaptation (and
many have already been explored and validated).
This stands in contrast to most prior work on natural language inspired programming and specification
(Section \ref{sec:relwork_natspec}), most of which has broadly either ignored linguistic treatments of natural
language semantics, or built on foundations that had not yet been thoroughly validated over a period of decades
(and ultimately did not stand up to such scrutiny).

This is especially important considering the three major ways a formalized, or controlled, fragment of natural language can fall short.
Any formal subset, as noted, omits some natural language out of necessity; the linguistics literature calls this \emph{syntactic under-generation}:
the formalized subset is a formal language, and undergeneration occurs when syntactically valid text is not accepted as syntactically valid.
There are two related issues, however. One is \emph{syntactic over-generation}: accepting text as syntactically valid that would not generally be accepted as valid
by humans. If such text can be written by mistake, and accepted, it is akin to an overly-permissive programming language parser: any interpretation is wrong,
as the text should have been rejected. The final major issue is misinterpretation: interpreting syntactically valid text in the formal subset in a way that
disagrees with reasonable human interpretations would result in a formal (controlled) natural language subset that is at best confusing, but more generally
significantly error-prone if used as a tool frontend.
While undergeneration is inherent to formalized natural language, work in linguistics on formal semantics and the interface of syntax and semantics is
evaluated largely on the absence of overgeneration and misinterpretation.
This is a key source in confidence of the accuracy of our formalization: both the core treatment and some extensions described in Section \ref{sec:add_grammar}
were read directly out of authoritative accounts in the linguistics literature.

  \section{Categorial Grammars for CIC Specifications}
  \label{open-categorial-grammars-over-cic}
  This section describes a model of a very small fragment of English for describing simple mathematics.
  Our goal is not to present a polished and complete natural language fragment suitable for a wide range of 
  nearly-arbitrary specifications; such a goal is admirable, but such large-scale projects take many
  years to complete~\cite{Marcus:1993,marcus1994penn,alshawi1992core}.
  Our goals are instead to demonstrate the feasibility of adapting extant linguistic characterizations 
  to natural language specifications; show that existing proof assistants (e.g., \lean) include enough 
  machinery to do so now, internally to the tools; and that the selected approach is modular enough to
   permit growing the scope of a system piecewise over an extended period of time.

  We describe how to use the typeclass support of modern proof assistants~\cite{sozeau2008first,selsam2020tabled} to perform semantic parsing from natural language\footnote{In this paper, English, but in principle any other natural language with thorough treatments in categorial grammar~\cite{hockenmaier2007ccgbank,hockenmaier2005ccgbank,hockenmaier2006creating,abzianidze2017parallel,ambati2018hindi,moot2015type,mineshima2016building}.} to the Calculus of Inductive Constructions (\textsc{CIC})~\cite{paulin1993inductive}, or alternatively (see Section \ref{sec:generalization}) embedded logics.
  This approach naturally supports an open-ended lexicon, which is essential to modularly expanding the set of words handled by semantic parsing 
  (Section \ref{sec:treebanks}).  
  This ensures the extended lexicon can grow in tandem with a formal development, 
  with organization chosen by developers rather than dictated by an external tool,
  and with the proof assistant automatically checking validity of 
  the lexicon extensions at the same time it checks validity of the rest of the formalization.

  While in principle the translation could be carried out via an external tool which may offer
  other benefits, this approach
  works in proof assistants today, without additional toolchain issues or concerns about disagreements
  between an external tool and a proof assistant over what is valid.

  Our goal is to allow verifiers to either (1) directly specify in English; or (2)
   carry out proof developments mostly as they already would,
  but to additionally state a specification in a formalized subset of English, and prove a lemma showing the traditional
  specification implies (or is equivalent to) the translated specification.
  Both cases result in a validataed, auditable connection between English and the proofs.
  
  We initially describe a general overview and notation we use throughout the paper. Section \ref{sec:perf} describes further changes
  which compromise readability but drastically improve parsing performance.

  \paragraph{The Basics}
  Most work on categorial grammars lumps all entities --- frogs, people, books --- into a single semantic type $E$ for ``entities'', 
  with different varieties distinguished by predicates: e.g., $\textsf{three} : E$ and $\textsf{Mary}:E$, but $\mathsf{number}(\textsf{Mary})=\textsf{false}$.  This approach is rooted in the assumption that first order logic is an appropriate semantic model for sentence meaning.
  That assumption is plausible for many general circumstances, popular in formal linguistics, and particularly useful for modeling figurative speech, 
  but stands at odds with making natural language claims about mathematical objects defined in intuitionistic type theory.
  For logical representations to talk about entities in \textsc{CIC} --- which distinguishes natural numbers, rings, monoids, and so on with different types --- adjustments must be made.
  Noun phrases and related syntactic categories must be parameterized by the semantic (\textsc{CIC}) type of entity they concern.

    We follow linguistically-motivated work using intuitionistic type theories like \textsc{CIC} for exploring possible logical forms~\cite{sundholm1986proof,ranta1991intuitionistic,ranta1994ttg,chatz2014nlcoq,bekki14dts,bekki2012LIRA,ranta1995context}, in using an alternative model where many grammatical categories are indexed by the underlying semantic type to which they refer. 

\begin{figure}
\begin{lstlisting}
inductive Cat : Type where
  | S -- Sentence/proposition
  | NP : forall {x:Type}, Cat
  | rSlash : Cat -> Cat -> Cat -- A/B
  | lSlash : Cat -> Cat -> Cat -- A\B
  | ADJ : forall {x:Type}, Cat
  | CN : forall {A:Type}, Cat.
\end{lstlisting}
\begin{lstlisting}
@[simp]
def interp (c:Cat) : Type :=
  match c with
  | S => Prop
  | @NP x => x
  | @ADJ x => x -> Prop
  | rslash a b => interp b -> interp a
  | lslash a b => interp a -> interp b
  | @CN x => x -> Prop
\end{lstlisting}
\caption{Core grammatical categories as an inductive type \textsf{Cat} and the mapping $\lfloor-\rfloor$ from grammatical types to semantic types as \textsf{interp}.}
\label{fig:corecoq}
\end{figure}
\textsc{CIC} is expressive enough to give the set of grammatical types as a datatype \lstinline|Cat|, and to give the interpretation of those types into semantic types as a recursive function \lstinline|interp| within \lean, as shown in Figure \ref{fig:corecoq}.
Grammatical types (called \emph{categories} in the linguistics literature) \textsf{Cat} include the aforementioned slash types, sentences ($S$); 
noun phrases ($\mathit{NP}~A$) denoting objects of type $A$ in \textsc{CIC}; adjectives ($\mathit{ADJ}~A$) 
denoting predicates over such objects\footnote{%
Not all grammatical treatments treat adjectives directly as primitive (some approaches derive them from other primitive categories), but all frameworks include
some for of adjective which is semantically a predicate.};
and common nouns ($\mathit{CN}~A$) denoting predicates on $A$, both refining the domain
of discourse and 
imposing constraints on the semantic type indices of other phrases in the context of a sentence, 
used in cases where a sentence must refer to a common class of objects (i.e., a type), such as ``natural numbers'' or ``rings.''  
As mentioned previously, the slash types correspond to function types (the direction is relevant only in the grammar, not the semantics), sentences are modeled by propositions (the type of logical claims), noun phrases of \lean type \texttt{t} correspond to elements of \texttt{t}, and similarly adjectives correspond to predicates on such types (i.e., elements of \lstinline|t -> Prop|).
These are modeled by the function \lstinline|interp|, which maps grammatical types to other \textsc{Lean} types.
Common nouns are interpreted as predicates as well --- this permits constructions such as ``even natural'' to be parsed as common nouns,
with semantics \lstinline-($\lambda$x:Nat, even x)-.
The \lstinline|@[simp]| definition marks the definition for unfolding during typeclass resolution.


  \label{a-categorial-grammar-dsl-in-coq}
\subsection{Combination Rules}
Parsing natural language specifications requires automatically applying rules like \textsc{/-Elim} to combine sentence fragments.
Rather than modifying a proof assistant, we can use existing trusted\footnote{Officially, typeclasses are not part of a trusted computing base, as they elaborate to record operations before being passed to the core proof checking apparatus. In practice, they mediate \emph{which} terms are passed to the core, so calling them \emph{untrusted} would be a misnomer~\cite{pollack1998believe}.} functionality to do this for us: typeclasses~\cite{sozeau2008first,selsam2020tabled}. These are a mechanism for parameterizing function definitions by a set of (often derivable) operations.  Proof assistants such as \coq and \lean permit declaring a typeclass (roughly, an interface), and declaring implementations associated with certain types.  The implementations may be parameterized by implementations for other types (such as defining an ordering on pairs in terms of orderings for each component of the pair).
  When a function is called that relies on a set of operations, the proof assistant attempts to use a form of higher-order unification to construct an appropriate implementation.
  It is possible to encode categorial grammar rules into typeclasses.
  Each judgment form corresponds to a typeclass, and each rule corresponds to an instance (implementation) of the typeclass.

  For now, assume we model the order of composition of sentence fragments 
  as a binary tree of words stored in leaf nodes, explicitly modeling the word groupings as
  bunched contexts in a substructural logic.
  Our actual implementation employs some optimizations (Section \ref{sec:perf}) which hide this structure and perform all work
  modulo associativity rules ($\Gamma,(\Delta,\Upsilon)\equiv(\Gamma,\Delta),\Upsilon$), but this
  version (consistent with an initial prototype) is a useful learning step.
 We will write \lstinline|#| to join two sequences of words. The constructor \lstinline|one| packages a word into a leaf node.

  We define the judgement form \(\Gamma\vdash T \Rightarrow e\) as:
  
\begin{lstlisting}
class Synth (ws:tree String) (c:Cat) where
  denotation : interp c
attribute [simp] Synth.denotation
\end{lstlisting}
If an instance of \lstinline|Synth ws C| exists, it comes with an operation \lstinline|denote| that produces a \lean value of type \lstinline|interp C| --- $\lfloor C\rfloor$.
 Because \(e\) is viewed as an output we would like
  to query, it is defined as a member of the typeclass, rather than as an
  additional index. When deriving a formal specification for sentence
  \(s\), we will arrange for the typeclass machinery to locate an instance
  of \texttt{Synth\ s\ S} --- checking that the sentence \(s\)  is a grammatically valid
  sentence --- and request its term denotation when necessary.
 Translating a specification given by the list of words \lstinline|ws| corresponds to parsing \lstinline|w| as a sentence: finding an instance of \lstinline|Synth ws S|.
 We define an instance for each rule to encode, 
  such as this one 
  corresponding to \textsc{\textbackslash-Elim}, which applies the logical form of the functional to the logical form of the argument:
\begin{lstlisting}
instance SynthLApp {s1 s2 c1 c2}[L:Synth s1 c1] [R:Synth s2 (c1 $\setminus$ c2)] : Synth (s1#s2) c2 where
  denotation := R.denotation L.denotation
\end{lstlisting}
and for leftward composition (\textsc{\textbackslash-Comp}):
\begin{lstlisting}
instance LComp {s s' c1 c2 c3}
  [L:Synth s (c1 $\textbackslash$ c2)] [R:Synth s' (c2 $\textbackslash$ c3)] 
  : Synth (s#s') (c1 $\textbackslash$ c3) where
  denotation x := R.denotation (L.denotation x)
\end{lstlisting}
The remaining rules of Figure \ref{fig:rules} can also be encoded in this way.
The \lstinline|[simp]| attribute on \lstinline|denotation| ensures it will automatically be simplified by tactics like \lstinline|simp| (roughly a combination of \coq's \lstinline|simpl| and parts of \lstinline|auto|).

In addition to exploiting the proof assistant's built-in search for parsing, the use of typeclasses means the set of rules is extensible,
meaning additional rules could be added for additional grammatical coverage or to speed up
the parsing process. More critically, however, it allows easy modular extension of the lexicon with additional words.

\subsection{Lexicon}
The lexicon is encoded via a typeclass \lstinline|lexicon|, which assigns grammatical types and semantics to individual words rather than series of words.
This is then tied to the \lstinline|Synth| typeclass:
\begin{lstlisting}
class lexicon (w:String) (c:Cat) where
  denotation : interp c
attribute [simp] Synth.denotation
instance SynthLex {w:String}{C:Cat}[lexicon w C] : Synth (one w) C where
  denotation := lexicon.denotation w
\end{lstlisting}
\lean{} permits declaring multiple instances for the same word (e.g., if a word has multiple meanings of different grammatical types), giving essentially a free variant of intersection types~\cite{moortgat1999constants} without the coherence issues described by Carpenter~\cite{carpenter1997type} (only one definition will be chosen per appearance of the word).
Thus, a dictionary for our approach consists of a set of instance declarations for \lstinline|lexicon|:
\begin{lstlisting}
instance fourlex : lexicon "four" (@NP Nat) where
  denotation := 4
instance evenlex : lexicon "even" (@ADJ Nat) where
  denotation := fun x => even x = true
instance n_is_adj {T}: lexicon "is" 
  (((@NP T) $\textbackslash$ S) / (@ADJ T)) where
  denotation := fun a n => a n
instance n_is_n_lex {T}: lexicon "is" 
  (((@NP T) $\textbackslash$ S) / (@NP T)) where
  denotation := fun a n => a = n
instance given_lex {A B}: lexicon "given" 
  ((@NP (A -> B)) $\textbackslash$ ((@NP B) / (@NP A))) where 
  denotation := fun f arg => f arg 
\end{lstlisting}
Here we have defined two different meanings for ``is'' allowing it to be used to apply an adjective (e.g., as in ``four is even''), or to denote equality (as in ``four is four'').  The difference between the two, beyond their denotation is the grammatical types: both expect a noun phrase to the left, and some other word to the right: an adjective in the first case, or another noun in the second.
Note that in both cases, the adjective or noun phrase must match the type of underlying \lean object the the left-side noun phrase refers to: the argument \lstinline|n| is in both cases a variable of type \lstinline|interp (@NP A)=A| because that is the argument of the outermost slash type, while the second argument in each entry corresponds to the interpretation of the second slash type's argument (a predicate or an additional term, respectively).
Coupled with a development-specific bit of lexicon to name a particular \lean object of interest:
\begin{lstlisting}[mathescape]
instance addonelex : lexicon Prop "addone" 
  (@NP (Nat -> Nat)) where 
  denotation := addone -- $\lambda$x. x + 1
\end{lstlisting}
this approach permits giving correct denotations to both:
\[\begin{array}{c}\llbracket\textrm{addone is monotone}\rrbracket \equiv \mathsf{monotone}(\texttt{addone})
\\
\llbracket\textrm{addone given 3 is 4}\rrbracket \equiv (\texttt{addone}\;3)=4
\end{array}\]

\subsubsection{Quantifiers}
\label{sec:quantifiers}
Quantifiers over $A$ can be given grammatical type 
\[\mathsf{Quant}~A \equiv  (S / (NP_A \backslash S) / CN_A) \]
abbreviated with a macro.
Thus, a quantifier looks to its right first for a common noun (corresponding to the word identifying the \lean 
type to quantify over), and after that is combined, the result looks further to the right for a sentence fragment 
expecting such a thing to its left. (After binding with the common noun, the remainder is in fact a continuation~\cite{barker2014continuations}.)
Then adding a lexicon entry for ``every'':
\begin{lstlisting}
instance forall_lex {A}: lexicon "every" (quant A) where
  denotation := fun N P => forall x, N x -> P x
\end{lstlisting}
and another for the common noun ``natural'' (number) allows correctly parsing sentences like
\[\llbracket\textrm{every natural is even}\rrbracket\equiv\forall (n:nat)\ldotp (\mathsf{even}\;n)\]
(Recall, we must still be able to state claims that are false.)
The common noun constrains the quantifier to work with noun phrases referring to natural numbers,
also using the predicate semantics to constrain the claim to those elements of the quantified type matching the predicate, as in\footnote{
  An additional grammar rule lifts adjectives to noun modifiers $CN/CN$, allowing ``odd natural'' to be parsed as a noun.
}
\[\llbracket\textrm{every odd natural is even}\rrbracket\equiv\forall (n:nat)\ldotp \mathsf{odd}\;n\rightarrow(\mathsf{even}\;n)\]

\subsubsection{Coordination}
\label{sec:coord}
One aspect of natural language which is the source of some interest is that the words ``and'' and ``or'' (or their equivalents in other languages) can often be used to combine sentences fragments of widely varying grammatical types.
For example, in ``four is even and positive'' the word ``and'' conjoins two adjectives: ``even'' and ``positive.''
Yet in ``four is even and is positive'' it conjoins two phrases of grammatical type $NP_\textsf{nat}\textrm{\textbackslash} S$ (``is even'' and ``is positive'').

We can directly adopt a solution from the computational linguistics literature~\cite{carpenter1997type}, and formalize that ``and'' and ``or'' apply to any semantic type that is a function into (a function into\ldots) the type \lstinline|Prop| of propositions.  
We define an additional typeclass to recognize such ``Prop-like'' grammatical types inductively, starting with the grammatical types $S$ and $ADJ$, and inductively including slash types whose result type is also ``Prop-like'', which define an operation to lift boolean semantics through repeated functions.
We then add a polymorphic lexicon entry for each of ``and'' and ``or'' which assigns them any ``Prop-like'' type.

Thus in a sentence like ``four is even and is positive'' the two conjuncts are recognized as Prop-like (their underlying semantic type is $\textsf{Nat}\rightarrow\mathsf{Prop}$), and the operations of the typeclass recognizing this automatically lift a binary operation on \lstinline|Prop| to a binary operation on predicates --- the classic pointwise lifting of the underlying Heyting algebra~\cite{lambek1988categorial}. For ``and'' this lifts logical conjunction to $\lambda P\ldotp\lambda Q\ldotp \lambda x\ldotp P~x\land Q~x$, which is exactly what is needed --- the grammar rules will apply this function to the semantics of the even and positive predicates, and finally 4.
Disjunction is handled similarly, and this generalizes to arbitrarily complex slash types whose final semantic result is \lstinline|Prop|.


\subsection{Using Specifications}
Defining a function from sentence representations to the denotations of the words in order is then relatively simple:
\begin{lstlisting}
@[simp]
def pspec (ws:tree String) [sem:Synth ws S] : Prop :=
  sem.denotation
\end{lstlisting}

When invoked with a tree of strings \lstinline|s|, \lean will search for an instance of \lstinline|Synth s S| --- a parse of the string tree as a complete sentence.  The semantics of a sentence has \lean type \lstinline|Prop| (a proposition, or logical claim, to be proven).
To complete the readable surface syntax from the introduction, rather than hand-constructing data structures, we define a macro \lstinline-[|...|]- that takes a sequence of words (technically, identifiers) and constructs 
the appropriate representation.
 
Thus we may translate a range of specifications given an appropriate lexicon, including those below (sugared into math notation for space and readability):

\[
\begin{array}{l}
\llbracket\textrm{addone is monotone}\rrbracket\\\qquad\equiv \forall x,y : \mathbf{N}\ldotp x\le y\Rightarrow \mathsf{addone}\;x\le\mathsf{addone}\;y\\
\llbracket\textrm{every natural is non-negative}\rrbracket\equiv \forall n:\mathbf{N}\ldotp n \ge 0\\
\end{array}
\]

\[
\begin{array}{l}
\llbracket\textrm{every natural is non-negative and some natural is even}\rrbracket\\\qquad \equiv(\forall n:\mathbf{N}\ldotp n \ge 0)\land(\exists n:\mathbf{N}\ldotp \mathsf{even}\;n)\\
\end{array}
\]

In particular, we can observe how these specifications manifest during interactive proof:
\begin{lstlisting}
def addone_mono : pspec [|addone is monotone|] :=
  by simp
     --⊢ ∀ (x y : Nat), x ≤ y → addone x ≤ addone y
     intro x y h ...
\end{lstlisting}
i.e., after simplification, a proof may continue either directly, or by appeal to a separate lemma stated purely formally in the case the proof's only goal is to bridge formal and natural language specifications.

If \lean cannot find a \lstinline|Synth| instance for a specification, the user sees one of two error messages
from \lean itself. One possibility is that \lean has exhaustively explored the possibilities and no parse exists
(this is typical of a specification using a word not in the lexicon).
The other possibility is that \lean has reached its timeout for typeclass instance search.
This is linear ``fuel''-type timeout parameter that may adjusted per-file, meaning that if a file using these natural 
language specifications requires longer search times, this can be done locally without forcing an entire development to use 
the longer search.
Anecdotally, use of this style of specification typically does require increasing this parameter, but on a
relatively old 4-core machine from 2015, most of the specifications discussed in this paper are parsed within the first 
author's reaction time, with Lean never exceeding 25\% of a single core during parsing, with 2.6GB of memory consumed. 
Sections \ref{sec:perf} and \ref{sec:case_study} address performance in more detail.
\looseness=-1

\subsection{Performance}
\label{sec:perf}
The performance of semantic parsing depends on both the underlying typeclass resolution procedure, 
as well as the space of derivations that must be explored during parsing (itself dependent on the data manipulated
within rules, as well as which grammatical rules are included).

The structural rules encoded in the \lstinline|Synth| typeclass instances are the primary drivers of search costs, along with the \lstinline|lexicon| 
instances.  Since most words have only one or a very small number of grammatical roles (in general, not just in our small 
prototype~\cite{hockenmaier2005ccgbank,hockenmaier2007ccgbank,hockenmaier2006creating}), we expect that lexicon ambiguity will \emph{not} be a major driver 
of search costs. Instead, most costs should arise from exploring 
the space of derivations.

The direct implementation approach described above becomes quite slow for sentences over a few words, so we apply a number
of optimizations, all well-established in the computational linguistics literature.

We exploit several classic optimizations from  work on parsing natural langauges with Prolog.
The first issue is that given a sentence of length $n$, there are $(2n!)/((n+1)!n!))$ ways to associate segments of that
sentence into a binary tree like that used in the earlier explanation (the $n$th Catalan number).
The na\"ive approach above requires exploring potentially all of these trees in order to yield a
 complete search procedure, which is prohibitively expensive for even modest
sentences. The solution is to represent the sentence parenthesization without explicitly manifesting unique structures.
There are established techniques for this in the literature on parsing via logic programming~\cite{pereira1987prolog,dahl1994natural}, specifically
the technique of \emph{difference lists}. The idea is that rather than representing the parenthesizations as a tree,
to represent it as a \emph{span}.  A difference list is a pair of lists $X$ and $Y$ where $Y$ is a \emph{suffix} of $X$, and the difference
list then represents the list segment from the start of $X$ to the suffix covered by $Y$ --- \lstinline|[3,4,5]| and \lstinline|[5]| represents the
prefix \lstinline|[3,4]|, but without explicitly constructing a fresh set of cons cells: the spine of the original list of words is reused.

Lean's use of tabled resolution~\cite{selsam2020tabled} (roughly, memoization) works nicely with this change in representation, but just as in
Prolog, memoizing based on large structures (specifically, lists of UTF-8 strings) is expensive. So we also apply the other standard parsing-via-logic-programming
optimization of representing lists not as explicit lists, but as natural numbers representing the starting index of the sublist (roughly, as how many elements
to drop from the list of words being parsed). 
So for a locally-fixed list \lstinline|["three","is","even"]|, rather than reprenting the full span as the list, or the explicit difference list pair of
\lstinline|["three","is","even"]| and \lstinline|[]|, the representation of the span is the pair of 0 and 3 (the list segment starting at index 0 and dropping index 3 and beyond).
Even in unary form this makes the table lookups faster (comparing unary naturals of maximum size/length $n$
is faster than comparing lists of maximum size/length $n$ containing strings which must be compared), but Lean additionally represents naturals
internally via the GMP library, making the comparisons even faster: naturals within range of a machine word are represented as a machine word (plus tag).
\looseness=-1

Applying this latter optimization unfortunately means we must also compute word-list indices using typeclasses, in order to tie lexicon entries
stated in terms of words to spans in terms of natural numbers. It also means that because the string is no longer an explicit part of the
context of the \lstinline|Synth| judgment/typeclass, we must use Lean's module boundaries to isolate specification searches from each other.
For each specification we open a new module, declare the string to parse to the typeclass machinery with a module-local instance, and then
use a variant of \lstinline|spec| adapted for the changes above. This adds a bit of verbosity, but could in principle be alleviated via
macros or additional facilities for exposing control over clearing the memoization tables; currently the only such mechanism for
this in \lean is that tables
are cleared of anything module-local when that module's checking is complete.
\looseness=-1

Finally, we apply standard techniques~\cite{eisner1996efficient} to further prune the search space of instances of spurious
ambiguity (searching only over normal form derivations), and impose two static bounds on the search: limiting how many times coordinators may undergo pointwise lifting,
and using \lean's existing heartbeat timeout counter, which gives roughly linear control over typeclass search time.


\paragraph{Worst Cases}
The particular grammar rules we are currently working with have only context-free recognition power, known to require
cubic time in the length of the sentence to parse.
The mildly-context-sensitive classes of categorial grammars favored by linguists have $O(n^6)$ worst-case parsing cost~\cite{joshi:91}, 
though in practice the common case can be made quite fast~\cite{Clark:2007,Clark:2003,Clark:2002}.

\section{Modularity and Extension: Growing a Lexicon, Handling More Logics}\label{sec:treebanks}
The previous section described only a small fragment of English suitable for formalizing mathematical claims.
Because categorial grammars are \emph{lexicalized} grammars (recall Section \ref{sec:formal_natural})
which use a small number of special-purpose rules (like those in Figure \ref{fig:rules}) and otherwise
leave knowledge of a language to per-word entries, they naturally support modular extension.
In particular, the
availability of slash types (directed function types) affords significant flexibility to define new grammatical roles without disrupting
the core rules,
and extensions to attach modalities to the slashes~\cite{Baldridge:2003:MCC:1067807.1067836,Moortgat1996} allow further constraints capturing the subtleties of natural language to be captured solely by giving precise grammatical types (and semantics) to individual words.

\subsection{Managing Words}
\label{sec:managing_words}
Adding new words to a categorial grammar lexicon is conceptually as simple as adding the word, particular grammatical type, and associated denotation to the database.  This makes it easy to extend a system with new concepts (e.g., new algebraic structures); lexicon entries to deal with concepts defined in a proof assistant library can be distributed as a part of that library.
Conversely, if a word or particular usage of a word is found to be confusing to humans, leading to ambiguity, or otherwise problematic, it can be removed from the lexicon while affecting only inputs that use that word in that way (i.e., the problematic ones).

In practice the situation will be more complex, but we expect most extension to require little, if any, special linguistic knowledge. Assuming a robust core lexicon, it is likely that most extensions will be additions of words with simpler categories.
Experiments on a large English lexicon showed~\cite{hockenmaier2005ccgbank} that when training on most of lexicon, the unseen words in a held-out test set were primarily nouns (35.1\%) or transformations of nouns (e.g., adjectives, at 29.1\%). These are the simplest categories to provide semantics for (types, objects, and predicates), strongly suggesting that proof assistant users with no special linguistics background could make most extensions themselves. Similar experiments for a wide-coverage lexicon of German~\cite{hockenmaier2006creating} show over half of unknown words to be nouns, suggesting this feasibility extends beyond just English.

Careful readers or prior students of linguistics may have wondered when matters of verb tense, noun case and number, grammatical gender,\footnote{Which does not exist in English, but does in German, French, and other languages} etc. would arise.  In full linguistic treatments, these are reflected in additional parameters to some grammatical categories.  So for example, in our setting a noun phrase would be parameterized not only by the underlying referent type, but also by the case, number and so on; lexicon entries would then carry these through appropriately (making it possible to for example, require the direct object of a verb to be in the accusative case rather than nominative). We have omitted such a treatment here partly because it would obscure the key ideas while adding little value, partly because many of these distinctions are less important for our examples in English (which has fewer syntactic case distinctions than other languages), and partly because some aspects (like tense) may make sense only for specific embedded specification logics.
We leave general-purpose treatments of these issues to future work.

\subsection{Supporting Additional Grammatical Constructions}
\label{sec:add_grammar}
Formalization of significant fragments of language much deal with more subtle constructions that what we have described so far can handle. However, what we have described thus far is essentially read directly out of the literature on linguistic semantics.  
Linguists have spent many decades building out knowledge of how to handle more sophisticated uses of quantification~\cite{steedman2012taking,moortgat1996generalized} (``every,'' ``some,'' ``most''), resolving pronoun references~\cite{jacobson1999towards}, discontinuity~\cite{morrill1995discontinuity} (where a word is far from a word it modifies), and much more~\cite{carpenter1997type,morrill2012type}.
Critically, because categorial grammars are lexicalized, \emph{most grammatical constructions require no special handling} given an appropriate base categorial grammar. 

Our experience thus far has borne out this claim of modularity.
As we have developed our prototype, we have only needed to modify or extend the core grammatical types of Figure \ref{fig:corecoq} for two reasons:

\paragraph{Prepositional Phrases}
We added a category for prepositional phrases, indexed by the variety of English preposition (\emph{of}, \emph{into}, etc.). The set of preposition indices is incomplete, but with the exception of constructions like ``\emph{of naturals}'' (which implies a need for a common noun, as opposed to ``\emph{of 3}''), these simply take a noun phrase to their right and have the identity function as semantics (i.e., ``\emph{of 3}'' simply denotes 3).
This is the standard treatment of prepositional phrases in compositional linguistic semantics.

\paragraph{Anaphora / References}
Some specifications will use indirect references, called anaphora --- pronouns (``it''), articles (``the''), and other 
words that have no self-contained meaning but instead refer to concepts used earlier in a sentence.
We have prototyped a refinement of Jacobson-style~\cite{jacobson1999towards} treatment of anaphora 
(references to things mentioned earlier in the sentence, such as pronouns or some uses of ``\emph{the}''); 
this involves the addition of another slash type $A\mid B$ for expressions of category $A$ if some kind of missing $B$ 
(i.e., what is being referred to) is resolved, and isolated the additional rules (which increase the size of the search space)
in a separate module (so those rules are not always on), including a variant for named variable references.
While there exist many categorial grammar solutions to the problem of anaphora~\cite{gerhard2005anaphora,morrill1995discontinuity,moortgat1996generalized,jacobson1999towards}, all of them rely on such an additional construction for sentence fragments with missing referents. Jacobson's approach is the basis for most
other treatments; unfortunately we are not aware of extensions of the normal form search pruning we use~\cite{eisner1996efficient}
to this feature, and consider such optimization future work.
\\

Beyond these changes, which are backwards-compatible with the exposition in Section \ref{open-categorial-grammars-over-cic}, we have added 
additional logical rules, such as lifting adjectives into noun modifiers (e.g.,. ``\emph{even}'' can be lifted for use as a modifier of 
``\emph{natural}''). Because these are presented as \lstinline|Synth| instances are modular additions to the core (they rely on some rules which slightly increase parsing time, so are in an optional module that need not be imported
unless pronouns are used).
Beyond that, all extensions are simply lexicon entries, notably all quantifiers, including uses of ``\emph{a}'' and ``\emph{any}'' 
embedded mid-sentence.

\subsection{Beyond \textsc{CIC}}
\label{sec:generalization}
While our framing so far has focused on generating specifications which in \lean{} have type \lstinline|Prop|, this is not required.
Categorial grammars require only that their top-level semantic truth type have the structure of a Heyting Algebra~\cite{lambek1988categorial}: 
a type with binary operators for standard logical operators.
\looseness=-1

Our Lean formalizations in fact makes this generalization: the core machinery is polymorphic over an arbitrary choice of 
Heyting Algebra, 
with a lexicon split between entries polymorphic over the Heyting Algebra being targeted (e.g., ``or'' and ``and'') and words
 specific to a given 
Heyting Algebra (e.g., an adjective given as a \lean predicate must target \lean's \lstinline|Prop|).

This means the core idea applies not only to specs of type Prop, but that this machinery can be readily retargeted to any logic formalized within the
proof assistant, such as LTL~\cite{pnueli1977temporal} or CTL~\cite{clarke1986automatic}. This is not itself novel
 (Section \ref{sec:relwork} 
discusses some prior approaches to this) but working directly within a formalized proof assistant brings accuracy 
benefits to such efforts. 
A partial formalization of Dzifcak et al.'s work~\cite{dzifcak2009and} in our original \coq prototype~\cite{nlspectr} revealed that they made use of invalid lexicon entries: 
in this paper's notation, entries for words with grammatical category $C$ whose specified semantics were not of type \lstinline|interp C|.

  \section{Trust and Auditing}\label{trust-and-auditing}
  
  One of the essential criteria for an LCF-style proof assistant is the
  production of an independently-checkable proof certificate~\cite{pollack1998believe}. While we
  have proposed using typeclass machinery to automatically parse and
  denote, and the typeclass resolution itself is typically not viewed as part of the trusted computing base (TCB), it does effectively produce a form of proof certificate.
  The typeclass machinery
  explicitly constructs an instance of the typeclass --- an element of
  the corresponding record type --- and passes it to \lstinline|pspec|.  
  So the proof assistant's kernel sees (effectively) a categorial grammar proof, constructed via typeclass instances rather than constructors of an inductive data type
  (with \lstinline|Synth| instance names in lieu of constructors).
  This explicit
  term persists into the compiled forms \lean already produces, and
  could be identified by an independent proof checker that wished to also
  validate the natural language interpretation.

  We can think of several ways a user might accidentally or maliciously risk confusing
  an independent checker.
  All but one can easily be detected by a checker aware of the categorial grammar specification typeclasses.
  The final possibility amounts to changing the specification in the proof certificate.
  \looseness=-1
  
  First, a user may redefine or extend our core instances (for \texttt{Synth}) to produce a different denotation.
    A certificate checker would already ensure these are type-correct. A natural-language-specification-aware extension could check that the \lstinline|Synth| instances correspond to the desired rules.
    Or to better support some of the extensibility arguments made earlier, the \lstinline|Synth| typeclass could be modified to also carry a 
    justification of its conclusions in a more general substructural logic~\cite{kruijff2000relating,gerhard2005anaphora}, which would amount 
    to requiring extensions to carry conservativity proofs over a trusted linguistic base system.
    \looseness=-1

    Second, a user may extend the lexicon with additional words or additional grammatical roles for a given word, introducing ambiguity into the parsing.  Checking for ambiguity is relatively straightforward: 
    setting aside  indexing by \cic types, equivalence of grammatical types is decidable, and a checker could conservatively require that any lexicon entries with the same index-erased grammatical types have clearly-distinct indices
    (in which case they would not unify during typeclass search under any circumstances).  An independent checker could verify the absence of ambiguity in the lexicon, or alternatively surface the use of any ambiguity in a parsing derivation for human inspection.

    Finally, a user could also manipulate the lexicon, for example \emph{redefining} (or \emph{mis}defining) ``monotone''
    to denote \(\lambda f, \mathsf{True}\). This is arguably a form of
    modifying the specification by changing definitions, rather than
    sneaking a broken proof past a certificate checker.  
    It is analagous to changing a definition of a property verified by a proof --- a working proof with the 
    wrong definition is wrong, but this leaves behind evidence of the incorrect definition, by leaving evidence of 
    how ``monotone'' was interpreted. This would however require human intervention to detect.

These possible forms of attack highlight the main sources of trust added when considering natural language specifications in the approach we describe: the grammatical rules for combining phrases, well-formedness of the lexicon, and the definitions of words in the lexicon.

Beyond these, there is the general issue of \emph{ambiguity} in parsing. Semantic parsing
gives rise to two forms of ambiguity: \emph{spurious} ambiguity (where there are
multiple parsing derivations, but they yield equivalent semantics), and \emph{true} ambiguity (where
the different derivations yield truly different semantics).
Spurious ambiguity is typically tackled by searching only for normal-form derivations~\cite{eisner1996efficient},
as we do.
Actual ambiguity can arise from multiple lexical entries with the same grammatical types but different meanings
as in the ``attack'' described above (for which we described mitigations), but can also
sometimes arise from matters such as quantifier scoping: in ``every child ate a pizza,'' was
there a pizza for each child (the most common reading), or was there a single pizza shared by all (less
common, but acceptable).
There are two approaches to mitigating these. First, there is linguistic evidence
for claiming that there truly is a most common resolution, and the grammar can be tailored to prefer
that linguistically more common result. For example, there is evidence that in English,
quantifiers earlier in a sentence (to the left) tend to outscope those to the right~\cite{barker2014continuations}, as in the example above.
Moreover, because of how quantifier nesting works formally, this is likely to be even further
emphasized in English for formal specifications.
Second, it is possible in general to perform an exhaustive search to identify ambiguity (then allowing
a rewrite to remove the ambiguity); \coq's typeclasses include options to perform exhaustive search to ensure
no ambiguity exists.


\section{Case Study: Sorting and Multisets in VFA}
\label{sec:case_study}
\begin{table*}[t!]
\caption{Translation experiments with \emph{Verified Functional Algorithms} specifications.}
\vspace{-1em}
\begin{tabular}{|c|c|l|}
\hline
\# & Ch & Original (formal) and Generated (if different)\\
\hline
\multirow{2}{*}{1} & \multirow{10}{*}{\rotatebox{90}{Sort}} & Original: ``insertion maintains sortedness'' (\lstinline|forall a l, sorted l -> sorted (insert a l)|)\\
& & Translated (6s): ``insertion of any natural maintains sortedness'' ($\alpha$-equivalent)\\
\cline{1-1}\cline{3-3}
\multirow{2}{*}{2} & & Original: ``insertion sort makes a list sorted'' (\lstinline|forall l, sorted (sort l)|)\\
& & Translated (5.32s): ``sort sorts any list of naturals'' ($\alpha$-equivalent)\\
\cline{1-1}\cline{3-3}
\multirow{2}{*}{3} & & [No Original English] (\lstinline|forall x l, Permutation (x::l) (insert x l)|)\\
& & Proposed \& Translated (7.4s): ``insert is a permutation of cons'' ($\alpha$-equivalent)\\
\cline{1-1}\cline{3-3}
{4} & & Original \& Translated (1.84s): ``sort is a permutation'' (\lstinline|forall l, Permutation l (sort l)|)\\
\cline{1-1}\cline{3-3}
\multirow{3}{*}{5} & & [No original English] (\lstinline|forall l, sorted (a l) /\ Permutation l (a l)|)\\
& & Proposed \& Translated (5.72s): ``sort is a sorting permuting algorithm''\\
&& (\lstinline|exists a, (forall l, sorted (a l)) /\ (forall l, Permutation l (a l)) /\ a=sort|)\\
\hline
{6} & \multirow{8}{*}{\rotatebox{90}{Multiset}} & Original \& Translated (1.27s): ``union is associative'' (\lstinline|forall a b c, union a (union b c)=union (union a b) c|) \\
\cline{1-1}\cline{3-3}
{7} &  & Original \& Translated (1.24s): ``union is commutative'' (\lstinline|forall a b, union a b = union b a|) \\
\cline{1-1}\cline{3-3}
\multirow{3}{*}{8} & & Original: ``insert produces the same contents as merely prepending the inserted element to the front of the list''\\
& & (\lstinline|forall x l, contents (insert x l) = contents (x :: l)|)\\
& & Translated${}^*$: ``insertion and cons of any value yield equal contents'' ($\alpha$-equivalent)\\
\cline{1-1}\cline{3-3}
9 & & Original \& Translated (1.27s): ``sort preserves contents'' (\lstinline|forall l, contents l = contents sort l|)\\
\cline{1-1}\cline{3-3}
\multirow{2}{*}{10} & & [No original English] (\lstinline|forall l, contents l = contents (sort l) /\ sorted (sort l)|)\\
& & Proposed \& Translated (7.18s): ``sort preserves contents and sorts'' ($\alpha$-equivalent)\\
\hline
\end{tabular}
\label{tbl:translations}
\end{table*}

Appel's \emph{Verified Functional Algorithms}~\cite{appel-vfa} is a textbook on verification of functional algorithms in 
\coq. To evaluate our prototype on non-hand-picked specifications, we translated formal lemma statements (but not proofs) of 
specifications from Chapters 2 and 3 of the book into \lean. 
These chapters deal with verification of insertion sort on lists of natural numbers, initially in terms of list properties
(Chapter 2) and then in terms of an extensional view of lists as an ordered multiset (Chapter 3).
Chapter 2 asks students to prove 5 specific lemmas about insertion sort and a helper function for insertion into a sorted list.
(It also includes lemmas proving equivalence of two definitions of a sorted predicate, but we omit these because internals of inductive definitions are not in the intended scope of the prototype.)
3 of these lemmas come with explicit English descriptions of the corresponding lemma statement, while 2 have no direct English correspondence given. 
Chapter 3 asks students to prove a range of specific lemmas about multisets, as well as additional specifications of
insertion sort in terms of multisets (where Chapter 2 uses the notion of list permutations).
We have translated 5 specifications from Chapter 2, and 5 specifications from Chapter 3, using our \lean implementation~\cite{artifact}.
For each of these we describe the original text (if given), a proposed adjustment with justification (if given), the original formalization, and the result of parsing the English into \lean with our prototype after adding appropriate lexicon entries. In cases where we changed the original text, we explain why. The purpose of this case study is to put pressure on the system regarding (1) its ability to express precise formal claims chosen without this system in mind, (2) its ability to (approximately) relate textbook-level English prose to formal specifications, and (3) to surface some linguistic issues at play in formal specification.
Table \ref{tbl:translations} summarizes our experiments.

In general, we reworded a number of examples which had explicit English translations in \emph{VFA},
most commonly to be more explicit about quantification, as in the first example. Even in this small sample,
English specifications tend to ellide details of what they quantify over. While it would be possible (due to our
categories being indexed by \cic types) to maintain this ellision, we opted to keep the explicit quantification
in part to minimize how much of English grammar our prototype required.
Our prototype does not contain a full formalization of the English
language (per the discussion of Section \ref{sec:formal_natural}), but none of the specifications we consider is beyond grammatical 
formalization: consider \textsc{CCGBank}~\cite{hockenmaier2007ccgbank}, which gives \textsc{CCG} categories and desired parses for 48,934 English sentences from
the Wall Street Journal, and is the basis of \textsc{CCG} grammars that have been used to parse even more complex texts, such as all of \emph{Alice in Wonderland}~\cite{yeung-kartsaklis-2021-ccg}.
Formalizing a more comprehensive English grammar is a long-term undertaking.
Our goal with this experiment was to explore both grammatical feasibility and implementation feasibility for smaller examples.

For Chapter 2, we have translated all five specifications from English into formal specifications that are
logically equivalent (for 4/5, $\alpha$-equivalent) to the hand-written specifications from the textbook.
We also proposed what we feel are reasonable English equivalents to specifications that were formalized but not
explicity described in English.
Example 5 is illustrative: it is clearly logically equivalent to the original, but because the English uses
the indefinite article ``a'' in a way that formal linguistic semantics typically treats as introducing an existential quantifier
(e.g., as in ``I have a duck''), the generated formal specification has an existential quantifier.
\looseness=-1

For Chapter 3, we have translated the specifications from the main portion of the chapter, omitting an extension treating
an alternative proof approach, where some specifications are complex strengthened inductive steps which are typically
not the sort one would specify in natural language.\footnote{Consider 
\lstinline|forall l x n, S n = contents l x -> exists l1 l2, l = l1 ++ x :: l2 /\\ contents (l1 ++ l2) x = n|\label{ftnote}}
\looseness=-1

The one specification we do not translate from that main section is the specification described in English by
``multisets in a nested union can be swapped.'' This text is both linguistically interesting, and under-specified.
It is ambiguous because, we believe, most readers would not understand what this meant from the text alone.
(Consider if you can understand it before checking the formalization in a footnote.\footnote{
  It is formalized as \lstinline|forall a b c, union a (union b c) = union b (union a c)|.
})
It is also linguistically interesting as it refers to the syntax used to write the property, rather than directly describing a property
of the union operation.
We could imagine a number of ways of making the exact English work in this case, or rewriting it slightly more generally
without recourse to syntax (e.g., defining ``un-nestable'' as a lexical entry and handling it similarly to 
examples 6 and 7), but we believe the proper way to address it is with a more thorough study of the linguistics of how 
mathematical text refers to syntax --- work which is a separate research agenda unto itself, building on what we have done here.

One specification, 8,
can be formalized, and the parse structure is not particularly complex, but \lean curently does not synthesize it
because at present it does not pull in the lexical entries for ``yield'' or ``equal'' during typeclass resolution,
for unknown reasons; the lexical categories are not more complex than other entries, and we can explicitly provide the entries
and obtain semantics $\alpha$-equivalent to the original, but \lean fails to produce a \lstinline|Synth|
instance for either of those 1-word spans of the sentence at a grammatical category matching the lexicon entry declaration.
We are actively investigating the cause of this.

\paragraph{Timing}
We are limited in the precision of our measurements for timing, as \lean exposes no direct way to measure
the search time for a particular typeclass resolution problem.
The times reported in Table \ref{tbl:translations} are the time to run \texttt{lake build}
(\lean's project build command) in the root of the project, after building the full project, then
removing all intermediate and final build artifacts for the particular specification, and running
\texttt{lake build} again to compile \emph{only} that file.
These time measurements then include process start-up time, time to parse the project file and
identify the file with the missing build, and to parse, typecheck (including the typeclass search), and
compile that \lean file.
Measurements were taken on a 2020 1.4GHz MacBook Pro with 16 GB of RAM.
This is enough to show the general range of times for parsing modest specifications; 
in a large proof development, we would not expect parsing of natural language style specifications
for top-level specifications (thus likely avoiding the complexities of finding language
for examples like that in footnote 12) to be a dominant cost.

\paragraph{Beyond Chapters 2 and 3}
We have also looked at later chapters of \emph{VFA} to anticipate other challenges, grammatical,
semantic, and implementation-related, which must be resolved in the long term.
The ``advanced'' portion of Chapter 3, which we did not formalize above, requires pronouns and textual variable names.
We have implemented grammatical support for named variables in text based on classic grammatical treatments
of pronoun binding, but extending the idea of normal form parsing to these models is unresolved in the computational
linguistics literature.
Multiple later chapters also shift from the monomorphic specifications of early chapters,
to polymorphic specifications. These introduce interesting open technical challenges on the categorial grammar
side. There, phrases such as ``every tree'' (as in binary search tree) is actually referring to \emph{three}
quantifications, not one: a quantification over key and value types, as well as a specific tree with those arbitrary key
and value types. This poses challenges for the decoding function \lstinline|interp|,
as well as for the grammatical constraints across the rest of a sentence.
Na\"ive extension of \lstinline|interp| runs into universe size issues: quantification of a type in \lstinline|Type n|
lives in \lstinline|Type (n+1)|. So straightforward approaches to extending \lstinline|interp| run into
problems where different cases should be returning types in different levels of the universe hierarchy.
This does not occur in traditional linguistics where all entities are of a single type $e$, or even in prior work
on type-theoretical semantics~\cite{chatzikyriakidis2017modern,ranta1994ttg}, where all types are explicitly
assumed to live in a single universe and polymorphism is not addressed (as the primary interest there generally remains
modeling general linguistic issues, not mathematical issues).
If that problem were resolved, there is then the problem that the indices of categories later in the sentence,
which would want to refer to trees with particular key-value types, cannot be directly indexed by types quantified
within the \emph{denotation} of another word. \lean has some support for declaring typeclass instances
with certain parameters held abstract, allowing \lstinline|Synth| instances to have arguments of the form
\lstinline|[forall T, Synth ... ((@NP T) \\ S)]|,
which could have the choice of \lstinline|T| supplied by another entry when computing denotations.

\section{Related Work}\label{related-work}
\label{sec:relwork}

\subsection{Categorial Grammars and Type Theory}
Categorial grammars and dependent type theories for natural language semantics have long histories~\cite{lambek1958mathematics,vanBenthem90,sundholm1986proof}. 
Our proposal differs from that work in our focus on building a system to describe dependent types directly in a system
where they are required, as opposed to most prior work's focus on using dependent types for linguistic outcomes.

Others have used type theories like \lean's for linguistic semantics~\cite{sundholm1986proof,ranta1991intuitionistic,ranta1994ttg,chatz2014nlcoq,bekki14dts}, broadly making the argument that variants of dependent type theory offer a range of appealing options for modeling natural language semantics,
and fix some percieved deficiencies in the use of a lambda calculus over first-order logic formulas. This work consistently focused on using this as a means to study linguistics.
The notion of indexing some grammatical categories by the type of a referent in such an underlying type theory comes from Ranta's work~\cite{ranta1995context} on studying the linguistics of mathematical statements, though this was focused purely on the study of mathematical language, and not on interfacing with mathematics work carried out in type theory.
Ranta~\cite{ranta1994ttg}, Kokke~\cite{kokke2015agda} and Kiselyov~\cite{kiselyov2015applicative} have formalized variants of categorial grammar with semantics in proof assistants, but only as object logics of study in order to prove properties of those systems, rather than as working parsers integrated with other uses of proof assistants. 
This leaves much to explore in integrating categorial grammar with various forms of type-theoretical language semantics~\cite{chatzikyriakidis2017modern}, some of which coincide with common specification patterns.

On the linguistic side, our particular choice to represent common nouns as \emph{typed predicates} appears to be new. 
Traditional linguistic semantics going back
to Montague~\cite{montagueEFL70} assume a single universe of discourse represented by a type \lstinline-e-, with nouns denoting predicates on \lstinline-e-
, so the noun \emph{even natural} would denote a predicate \lstinline|($\lambda$x, even x $\land$ natural x)|). Traditional type-theoretic semantics treats common nouns exclusively as
types~\cite{chatzikyriakidis2017interpretation,ranta1994ttg} (an even natural there would be a subset type). 
Retor\'e~\cite{retore2013montagovian} treats common nouns as \emph{either} types or predicates
on a single universe of discourse, depending on the circumstance. We treat them as both \emph{simultaneously}, allowing them to play a role in
unification during parsing while also contributing further refining behaviors. We believe that this is likely to work better for a wide variety
of formalizations, as subset types remain relatively less-frequently-used in proof assistants; we leave validating this conjecture to
future work.

\subsection{Natural Language for (Semi)Formal Specifications}
\label{sec:relwork_natspec}
We are hardly the first to argue for narrowing the gap between natural language and formal specifications,
Nor are we the first to attempt this via formal grammars that also model semantics.

Seki et al.~\cite{seki1988processing,seki1992method} is the earliest approach we are aware of, using an alternative lexicalized grammar formalism 
\\(HPSG~\cite{pollard1994head}) to translate natural langauge to first-order logic.
Their prototype was divorced from any particular use of the resulting formal specifications.

Dzifcak et al.~\cite{dzifcak2009and} used \textsc{CCG}s to translate natural language specifications to $\textrm{CTL}^*$, though as mentioned earlier
their semantics contain semantic type errors which are caught by working within a proof assistant that \emph{enforces} consistency between grammatical and semantic types
 (i.e., caught by the dependent type of \lstinline|Synth.denotation|).
They also translate to PDL, reusing some entries by not type-checking their semantics, enabling the mistakes above.

One recurring theme in formalized natural language interfaces to logics is  intentionally (by design) reflecting aspects of the syntax of a target
logical language back into their handling of natural language syntax,
in a way that superficially \emph{appears} to be consistent with the natural language, but in fact leads to subtly incorrect semantics. 
For example, \textsc{PENG}~\cite{schwitter2002english} and its derivatives
reflect the operator precedence of first-order logic into its interpretation of English, which is wrong, leading to misinterpreting
phrases like ``not yellow or blue'' as in ``the signal is not yellow or blue'' as $(\neg\mathit{yellow})\vee\mathit{blue}$ because negation binds more tightly
than disjunction \emph{even in \textsc{PENG}'s English surface syntax}, when most English speakers
would interpret the phrase as $\neg(\mathsf{yellow}\vee\mathit{blue})$.
Focusing on actual linguistic treatments of syntax and semantics as we do leaves such phrases ambiguous; while we have not implemented ambiguity 
detection,
ambiguities cannot be caught if they are intentionally not modeled. While we have set aside our \coq prototype~\cite{nlspectr}, 
\coq's typeclass implementation has a flag
\texttt{Typeclasses Unique Solutions} which verifies there is exactly one instance (e.g., of \lstinline|Synth|) solving a problem, and fails otherwise; such a feature
could be added to \lean and would then implement ambiguity detection for our encoding (in addition to its other uses for typeclasses in general).\footnote{This
relies on our use of normalized parse trees.}
\textsc{PENG} also restricts use of adjectives to a single adjective immediately before a noun, while our grammar is more general
without much active effort (beyond choosing an established, known-generalizable grammatical and semantic basis):
 the rule lifting an adjective to a common noun modifier ($CN/CN$) applies, and then handling of phrases like
``positive even prime natural'' simply fall out of \textsc{/-Comp} and \textsc{/-Elim}.
\looseness=-1

The most successful and long-lived prior effort in this space (in active development for over 20 years), by most metrics, is \emph{Attempto Controlled English} (\textsc{ACE})~\cite{fuchs1999controlled},
which is also emblematic of the philosophy behind most controlled natural language~\cite{fuchs2009controlled,kuhn2014survey}.
\textsc{ACE} is a highly-regimented formal fragment of English~\cite{ace_syntax} which aims to be an \emph{editorialized} version of English grammar 
meant specifically for first-order logic specifications.
Its primary implementation uses unification-based parsing, as in our work and most categorial grammar work, via Prolog's definite clause grammars~\cite{pereira1980definite,pereira1987prolog}.
\textsc{ACE} is widely credited as one of the earliest attempts to specify software in English~\cite{fuchs1999controlled,fuchs1996ace}, by
translating specifications into Prolog clauses, which can then be used as a knowledge base
for queries. This kind of controlled natural langue interface is a classic application of Prolog, predating \textsc{ACE}~\cite{warren-pereira-1982-efficient}. 
However, this results in a system that is disconnected from toolsets that people have ended up commonly using to specify software
implementations: using an \textsc{ACE}-translated specification in another tool would require additional engineering work to export the Prolog representation of
the specification to a form used by another tool (e.g., Z3). While not a fundamental issue, to the best of our knowledge, this engineering work has never occurred. 
\textsc{ACE} and similar systems also attempt to reduce the burden of growing a lexicon by treating most words as essentially uninterpreted functions or relations,
rather than tying into existing formal definitions.

A significant philosophical difference between most controled natural language systems and our goals is the designer expectations for how
restrictive the grammar of a system might be. \textsc{ACE}, \textsc{PENG}, and other systems generally assume the set of grammatical categories is fixed
at system design time, forever. Per our discussion in Section \ref{sec:managing_words}, as most lexical entries have obvious basic categories such as nouns
and adjectives, adding additional lexical entries in those categories, as in our work, is straightforward.
However, adding new grammatical constructions in these earlier frameworks requires adding new non-terminals to their grammars.
Because most linguistics research of the past several decades does not use standard phrase structure grammars (\textsc{CFG}s),
this leaves such extensions in a difficult position: implementors must either roll their own extensions in a way that has never been validated by
linguists, or are stuck with grammatical treatments that are decades out of date.
In contrast, our experience in Section \ref{sec:add_grammar} shows that the linguistics literature on categorial grammars offers 
modular grammatical extensions to new features which have been extensively-vetted by linguists, and can often be directly transcribed from the linguistics
literature. This is in addition to the fact that in many cases these extensions do not require modifying core rules, but
instead follow from additions of individual lexicon entries (because categorial grammars are lexicalized).

We believe there are two primary reasons that the work above has not lead to widely-used controlled natural language specifications, both of which we avoid.
The first is that prior implementations have been designed to prioritize parsing and translation over use of the resulting specifications;
most prior work produces Prolog representations of specifications. This is primarily an engineering problem, but of the sort that frequently hampers adoption
of ideas. Our implementation is a library for an actively developed proof assistant that is actively used for specification and verification of mathematical claims
about both software and mathematics, using features that already exist for purposes unrelated to natural language interfaces.
\looseness=-1

The second is that prior work made what we believe are suboptimal choices regarding which influences from linguistics to incorporate. These include conscious
choices like \textsc{PENG}'s aforementioned reflection of FOL operator precedence into English (similar confusion between natural English interpretation and how words are repurposed
for specific logics also lead to confusions with LTL~\cite{greenman2022ltl}).
It also includes cases like Vadera and Meziane's work~\cite{vadera1994english}, where not only were similar unnatural heuristics reflected back into English (so users had
to remember, and be able to consistently apply, the implementation's heuristics for quantification in order to understand specifications), but time has
since revealed the work they were built on to be far less complete treatments of the linguistic phenomena than were claimed (in good faith) at the time.
Vadera and Meziane adopted Hess's~\cite{hess-1985-natural} approach to resolving ambiguities in quantifier scoping
(an approach motivated by dismissing as mathematically impossible approaches that not only worked mathematically, but are now accepted as standard, thoroughly empirically validated
treatments~\cite{szabolcsi1997ways,szabolcsi2010quantification}).
These types of issues are why we have taken pains to work with an approach to controlled natural language that borrows as heavily as possible from
widely-vetted linguistic theories: while it is relatively straightforward (though significant work) to choose a syntactic subset of a natural language,
without specific efforts to draw on linguistic theories of meaning with significant empirical validation (and in particular validation of combined 
syntax and semantics)
it is easy to accidentally (or intentionally) formalize a controlled natural language in such a way that the natural and formal interpretations of text 
silently diverge in critical ways, or unnecessarily restrict the language to avoid those issues.

This is the key limitation of what may be the most closely-related work to what we propose here: two systems that focus on controlled natural language in proof assistants.
The earlier is work on GF-Alfa~\cite{hallgren2000extensible}, which used Ranta's Grammatical Framework (GF)~\cite{ranta2004grammatical} to add translation from English
to a proof assistant based on Martin-L\"of type theory. GF is a toolkit for building bidirectional relationships between logical form and text: for both parsing, and generation (discussed in the next subsection).
The paper describes the implementation of writing natural language specifications, but does not discuss actual use of the feature.
The other system is current, developed contemporaneously with our earlier \coq prototype: Naproche~\cite{de2021isabelle} is an integration into the Isabelle proof assistant of the ForTheL controlled natural language~\cite{forthel}.
ForTheL is a substantial template-based controlled natural language. While significant amounts of undergraduate algebra and set theory have been specified in ForTheL,\footnote{\url{https://github.com/naproche/FLib}}
the language is centered around describing new definitions (which then act as noun phrases), and \emph{notions}, which are predicates usable as common nouns or adjectives.
It is not possible to add additional verbs, so half of our case study sentences (1, 2, 8, 9, and 10) cannot be specified in ForTheL without significant rewriting 
 to avoid
non-standard verbs; this applies to many other \textsc{VFA} specifications as well.
In any case, because both systems use ad hoc grammars that are not based on established linguistic theories, extensions could easily go awry by over-generating
or misinterpreting. This is slightly less likely with ForTheL specifications than with GF-based specifications: ForTheL's grammar is quite restrictive, while GF
grammar definitions encourage regular expression style descriptions of arbitrary text fragments, which can easily lead to unexpected parses.
These systems, unlike ours, do permit associating definitions and proofs with controlled English text (both involve changes to the proof assistant interfaces specifically for controlled natural language interfaces), with the same caveats about grammatical expressivity.
We suspect we could also support definitions by additionally using \lean's macro facilities (which remain usable by libraries, without modifying \lean) to register the definitions, but believe this is actually counter-productive
for type-theory-based systems, as the exact form of an inductive type and computational behavior of function definitions can have major impacts on the difficulty
of a proof; as earlier, we recommend instead proving relationships between manual formal definitions and descriptions translated from English.
Neither of these systems produce anything akin to a proof certificate recording how English text was translated.

Our use of categorial grammar is in some ways also a hedge against the possibility that the relevant linguistic theories are further revised or refined;
categorial grammars are naturally open-ended and modular, allowing further refinements to be overridden.
Picking a basis with extensive linguistic validation also opens the door, in the future, to exploiting linguistically-grounded practices in
grammar engineering~\cite{bender2021grammar,ranta2011grammatical} and \emph{semantics-aware} grammar induction~\cite{KwiatkowksiZGS10,Kwiatkowski2011} to ease growing
the formalized coverage of the natural language over time --- a path not available to attempts not rooted in extensive linguistic coverage.

Seki et al.'s early work using \textsc{HPSG}~\cite{seki1988processing,seki1992method}, turns out to be prescient: \textsc{HPSG}
is also a lexicalized grammar formalism, and is now roughly as thoroughly investigated as categorial grammars~\cite{muller2021head};
it would be reasonable to pursue our agenda based on \textsc{HPSG} rather than categorial grammar, though we believe categorial grammar's connections
to substructural logic make it a slightly better basis for proof assistant users. 
However, at the time, \textsc{HPSG} was nascent: it was first proposed only 3 years before Seki's initial attempt~\cite{seki1988processing},
and only widely publicized in the year prior to Seki's work~\cite{pollard1987information}.
So at the time, \textsc{HPSG} did not offer the extensive library of formal treatments of linguistic phenomena that are now available for
both categorial grammars and \textsc{HPSG}.
\looseness=-1


More recent closely related is work on using natural language to describe \emph{tests}. Most recently, in parallel with this work, Gordon~\cite{icsenier22} proposed using categorial grammars to generate property-based tests~\cite{claessen00quickcheck} from English, using off-the-shelf \textsc{CCG} tools to generate an abstraction of JavaScript tests. While using \textsc{CCG}s for parsing in principle makes Gordon's approach extensible like ours, that work's lexicon is sufficiently limited that the English accepted is fairly formulaic and template-like.
Harris and Harris~\cite{harris2015generating} used a variant of context-free grammars to generate hardware tests expressed in \textsc{CTL}, realizing an idea first proposed (but never evaluated) by Nelken and Francez~\cite{nelken1996automatic}.

\paragraph{From Formal Specifications to Natural Language}
More distantly related is work translating the reverse direction from formal specifications to English~\cite{johannisson2007natural,burke2005translating},
and work on expressing proofs themselves in (semi-)natural language~\cite{wenzel1999isar,cramer2009naproche}, though these all emphasize
highly-restricted fragments of natural language, rather than using foundations that capture language structure more generally.
In general, categorial grammars can be used for generation of text from a logical representation, intuitively by running the parsing unification process ``in reverse,''
searching for a parse tree whose semantics are equivalent to the logical form being described~\cite{kay-1996-chart}.
This approach has the advantage of being able to use the same grammar (combination rules and lexicon) that are used for parsing, so in principle
a grammar grown for parsing as we have proposed in this paper can be repurposed. 
This has been seriously-explored for \textsc{CCG}s in particular~\cite{white-baldridge-2003-adapting,white-2006-ccg,white-etal-2007-towards,Nakatsu_White_2010}.
But in practice, this approach to text generation is more difficult than semantic parsing, because the search space 
(all derivations whose semantics match a target) is much larger, and these approaches do not currently handle quantifiers, which are essential for formal specifications.
Additional challenges arise in taking human-written formal specifications as input, as the grammar rules yield unreduced function applications
(so at least $\beta\eta$-equivalence would need to be considered),
and some semantic treatments related to quantifiers would require considering full logical equivalence in the target logic (consider the existential in Sentence 7 of Table \ref{tbl:translations}).
\looseness=-1


Existing tools for translating from formal specifications to English
avoid the difficulties just highlighted by either restricting the specifications they handle (e.g., specification languages without quantifiers~\cite{burke2005translating,johannisson2007natural}),
or using more ad hoc treatments of grammar (as in Ranta's GF, which essentially translates first-order logic to English via a recursive function from formula syntax trees to strings,
filling in a template for each node type~\cite{ranta2011translating}).
GFT was previously used in the aforementioned GF-Alfa system~\cite{hallgren2000extensible} to translate formal specifications into
English. The case study reported there suggests that the workflow involved heavy specification-specific customization of the grammar, in a way supported by
user interface extensions in GF-Alfa, but seemingly requiring significant additional manual effort for each specification translated to English.

\subsection{Controlled Natural Language Beyond Specifications}
Also related is the long history of work on \emph{programming} in (controlled) natural language, including both program text itself~\cite{ballard1979nlc,cook2007applescript} and
interactions with IDEs~\cite{price2000naturaljava} manipulating programs written in standard programming languages (e.g., Java).
Most prominently, \textsc{COBOL} was intended to make code ``readable by managers or other non-programmers''~\cite{schneiderman1985relationship}.
Most attempts at programming languages in a natural language style were targeted at non-technical or less-technical users, and at
specific domains (such as Applescript~\cite{cook2007applescript}), even if the targeted language was actually quite general (e.g., \textsc{COBOL}~\cite{schneiderman1985relationship}).
None of these systems were originally intended to have user-extensible grammars, though some, including \textsc{COBOL}, were later supplemented with
token-based macro preprocessors~\cite{volpano1984empirical}, which have modularity problems now well-known from experience with the \textsc{C} preprocessor.
Some of these, notably \textsc{COBOL} and \textsc{Applescript}, have seen significant use for extended periods of time by many users in their target domains.
Attempts at fully-general natural language programming~\cite{ballard1979nlc} have been less successful, largely because they lead
to cases where the natural language is interpreted unexpectedly~\cite{miller1981natural,biermann1983experimental}, \emph{and there is no way within the system to correct this misinterpretation}. In particular,
we can find no prior system for general-purpose natural language programming which includes a fall-back to a more traditional (and less ambiguous)
programming language (most efforts were aimed at non-programmers).
Our proposal avoids this problem: we are proposing to \emph{supplement} a deeply formal system for specification and proof with controlled natural language,
not supplant the formal approaches. The intended users of our approach are experts in formal specification who wish to record the relationship between
(controlled) natural language and formal specifications. Incompleteness or unexpected behaviors of our approach (which we have already argued can be modularly
adjusted in general) will never prevent a \lean user from completing their proof, but only prevent them from recording the formal-natural relationship
in a rigorous way.

\subsection{Autoformalization}
What we do in this paper falls under a broad interpretation of what is now known as \emph{autoformalization} in the machine learning community. Current techniques in this space have promising results, with one system~\cite{llm-autoformalization} successfully formalizing \emph{and proving} (in Isabelle) 35.2\% of a collection of English Math Olympiad-style problems.
While still impressive (there is no explicit lexicon), this statistic ellides important limitations, including those fundamental to large language models (mentioned earlier), and specifics of the datasets used (which extended to pre-university math problems, but also includes significant primary-school-level math).
Most critically, machine learning approaches currently used for this are built on techniques
which consistently struggle to correctly handle even slightly-nontrivial boolean reasoning~\cite{ettinger2020bert,pandia-ettinger-2021-sorting,traylor2021and,truong-etal-2023-language}
(let alone more complex logical operations like universal quantification~\cite{asher-etal-2023-limits}).

\section{Looking Forward}
We have presented evidence that it is plausible to support natural language specifications in current 
proof assistants by exploiting existing typeclass machinery, with no additional tooling required.  
Carried further, this could be useful in many ways.  It can reduce the gap between informal and formal 
specifications, reducing (though not eliminating) trust in the manual formalization of requirements.  
Potentially non-experts in verification could understand some theorem statements, gaining confidence 
that a verification result matched their understanding of desired properties. And this could be used in 
educational contexts to help students learn or check informal-to-formal translations.

Of course, the details matter as well, and it will take time to realize a prototype that is broadly useful. 
First and foremost, a rich lexicon is required. As explained earlier, at least the initial lexicon will 
need to be manually constructed (borrowing grammatical categories from existing lexicons~\cite{hockenmaier2007ccgbank,abzianidze2017parallel}, and filling in the 
semantics) before it would be fruitful to adapt techniques for learning lexicons~\cite{Artzi:16spf,kanazawa1995learnable,Kwiatkowski2011} to extend
the manually-crafted base. Guiding this effort would 
require a substantial collection of examples of natural-language descriptions of formal claims, both for 
prioritizing lexicon growth and for validation that the approach is growing to encompass real direct 
descriptions of claims. 

Our performance results, while modest in scope, are informative, highlighting that many specifications
are likely to be parseable in times that are tolerable in interactive settings. External implementations
of parsers could potentially be used to speed up parsing, and they could emit proof certificates in the
form of explicit construction of our \lstinline|Synth| instances. However this requires additional
layers of separate tooling from the typeclass-based approach explored here, and also risks
problems with incompatibilities between tools and specific proof assistants; working within
the proof assistant guarantees all generated semantics are well-typed (even in the presence
of features like implicit arguments and rich macro support in defining semantic terms),
a problem which has arisen in other settings (as in our discovery that some of Dzifcak's lexical
entries have type-incompatible semantics~\cite{dzifcak2009and}, or as in Gordon's report~\cite{icsenier22}
that NLTK's combination rules sometimes introduce type errors into semantics).
\looseness=-1

It is possible that small differences will be required between standard natural language grammars and 
those used by this approach, arising from distinctions important to proof assistants but irrelevant to 
colloquial language. This is already the case, as mentioned, with the indexing of some grammatical 
categories with the semantic types of referents, following Ranta's early work on formalizing mathematical
 prose~\cite{ranta1995context}.
This direction offers opportunities to collaborate with linguists working in syntax and compositional
 semantics~\cite{barker2007direct,jacobson2014compositional,steedman2012taking}. 
Such collaborations could both help with possible novel linguistic features of ``semi-formal'' natural 
language, and offers a setting for applying classical linguistic techniques in a domain where they provide 
unique value.

A great deal of work lies ahead, but the potential benefits seem to more than justify further exploration in this direction.

\begin{acks}
This work was supported in part by the \grantsponsor{NSF}{US National Science Foundation}{https://www.nsf.gov/} under
Grant No.:~\grantnum[https://www.nsf.gov/awardsearch/showAward?AWD_ID=2220991]{NSF}{CCF-2220991}.
\end{acks}

\bibliography{csg,vmm,ccg,effects,other,education}

\end{document}